\documentclass[12pt,preprint]{aastex} 
\usepackage{psfig}

\newcommand{\et}{et al.}

\newcommand{\kms}{\mbox{km s$^{-1}~$}}

\begin{document}  
  
\title{Optical SETI: A Spectroscopic Search for Laser Emission \\  
    from Nearby Stars$~^{1}$}  
  
\author{Amy E. Reines\altaffilmark{2} and  
    Geoffrey W. Marcy\altaffilmark{2,3}}  
  
\email{gmarcy@etoile.berkeley.edu}  
  
\altaffiltext{1}{Based on observations obtained at the  
W.M. Keck Observatory, which is operated jointly by the  
University of California and the California Institute of Technology.}  
  
\altaffiltext{2}{Department of Physics and Astronomy, San Francisco   
    State University, San Francisco, CA USA 94132}  
  
\altaffiltext{3}{Department of Astronomy, University of California,  
Berkeley, CA USA 94720}

\begin{abstract}  
We have searched for nonastrophysical emission lines in the optical
spectra of 577 nearby F, G, K, and M main--sequence stars.  Emission
lines of astrophysical origin would also have been detected, 
such as from a time--variable chromosphere or infalling comets.  We
examined $\sim$20 spectra per star obtained during four years with the
Keck/HIRES spectrometer at a resolution of 5 \kms, with a detection
threshold 3\% of the continuum flux level.  We searched each
spectrum from 4000~\AA~-- 5000~\AA~  
for emission lines having widths too narrow to be natural
from the host star, as well as for lines broadened by
astrophysical mechanisms.  We would have detected lasers that emit a
power, $P>$50 kW, for a typical beam width of $\sim$0.01 arcsec
(diffraction--limit from a 10-m aperture) if directed toward Earth
from the star.  No lines consistent with laser emission were found.
\end{abstract}  
  
\received{} 
\accepted{} 
\slugcomment{Submitted to Pub.Ast.Soc.Pac.}  
  
\section{Introduction}   
   Most searches for signals from extraterrestrial intelligent
civilizations (SETI) have concentrated on radio wavelengths
(e.g. Tarter 2001, Cullers 2000, Werthimer et al. 2000, Leigh and
Horowitz 2000).  However, a preferred wavelength for SETI work has
proved elusive.  The motivation for this project stemmed from the
suggestion of Schwartz and Townes (1961) that advanced civilizations
might use optical or near--infrared lasers as an efficient means of
interstellar communication (Townes 1998, Lampton 2000).  The
advantages of optical wavelengths over radio include the higher
bandwidth and the narrower diffraction--limited beam size to promote
private communication.  In addition, the technology needed to produce
lasers with detectable powers is not far from current Earth technology
(\S 2.3).

Optical SETI (OSETI) projects are currently designed to detect highly
energetic, pulsed lasers and continuous--wave lasers. Such efforts
include dedicated searches, piggyback searches, and data mining.  Many
OSETI projects are designed to detect pulsed lasers using fast
photometry (Howard \et~2000, Bhathal 2000, Werthimer \et~2001).  Two
or three photodetectors are employed to search for laser pulses having
a duration of $\sim$1 ns in broadband visible light.  A high--powered
nanosecond pulsed laser will outshine its host star during its brief
pulse because at most one photon will be received (for a typical 1-m
class telescope) from the star during the 1 ns duration.

Alternately, interstellar optical communication may occur via
nearly continuous laser emission.  In that case, lasers of even
moderate power, tens of kilowatts, can outshine the host star 
when focussed by large apertures during transmission (see \S 2.3).  
Here we report a search for ultra narrow emission
lines such as predicted by the Schwartz--Townes suggestion.

\section{Optical SETI with High Resolution Spectroscopy}  
   We examine high resolution spectra of 577 F, G, K, and M
stars,   most of which reside within 50 pc.  We have taken
$\sim$12,500   spectra using the HIRES echelle spectrometer on the
10--m Keck 1 telescope.  All target stars are listed in
Nidever et al. (2002).  The stellar spectra were obtained during the
period 1997--2001 in a search for extrasolar planets using
relative radial   velocities.  We are indebted to Drs.~R.~P.~Butler
and S.~S.~Vogt for their contributions in obtaining these Keck
spectra.  The multiple spectra from 4000~\AA~-- 5000~\AA~ 
are compared to detect sharp differences such as those caused by
nonnatural emission lines, as described in the following subsections.  

\subsection{High Resolution Echelle Spectra}  
  
The acquired echelle spectra contain 34 spectral orders, covering  
wavelengths 3800 \AA -- 6100~\AA~with resolution $\lambda / \Delta  
\lambda$ = 70,000 (see Vogt et al. 1994, Vogt et al. 2000, Butler et  
al. 2000).  Not all orders are used here because the 
starlight is sent through a glass cell  
filled with iodine vapor before entering the spectrometer for use
in setting the wavelength scale in the planet search.
The iodine  lines contaminate the stellar spectrum at wavelengths
greater than 4980 \AA.  The remaining  
usable orders, 72 through 89, contain 1000~\AA~ranging from 3980~\AA~--  
4980~\AA.  This region includes Balmer $\beta$ and $\gamma$, but not
the CaII H\&K lines.   The pixel width is given by  
$\lambda/ \Delta \lambda$ = 145,000 , 
corresponding to 2.1 \kms per pixel or $\sim$0.032~\AA~per pixel.
The actual spectral resolution is only 70,000 because the
instrumental profile has a width of 2.3 pixels.
  
\begin{figure}  
%%\centerline{\scalebox{.75}{\rotatebox{90}{\includegraphics{fig1.ps}}}}  
%\psfig{figure=fig1.ps,angle=90,width=\textwidth}
%\psfig{figure=fig1.ps,angle=90}
\plotone{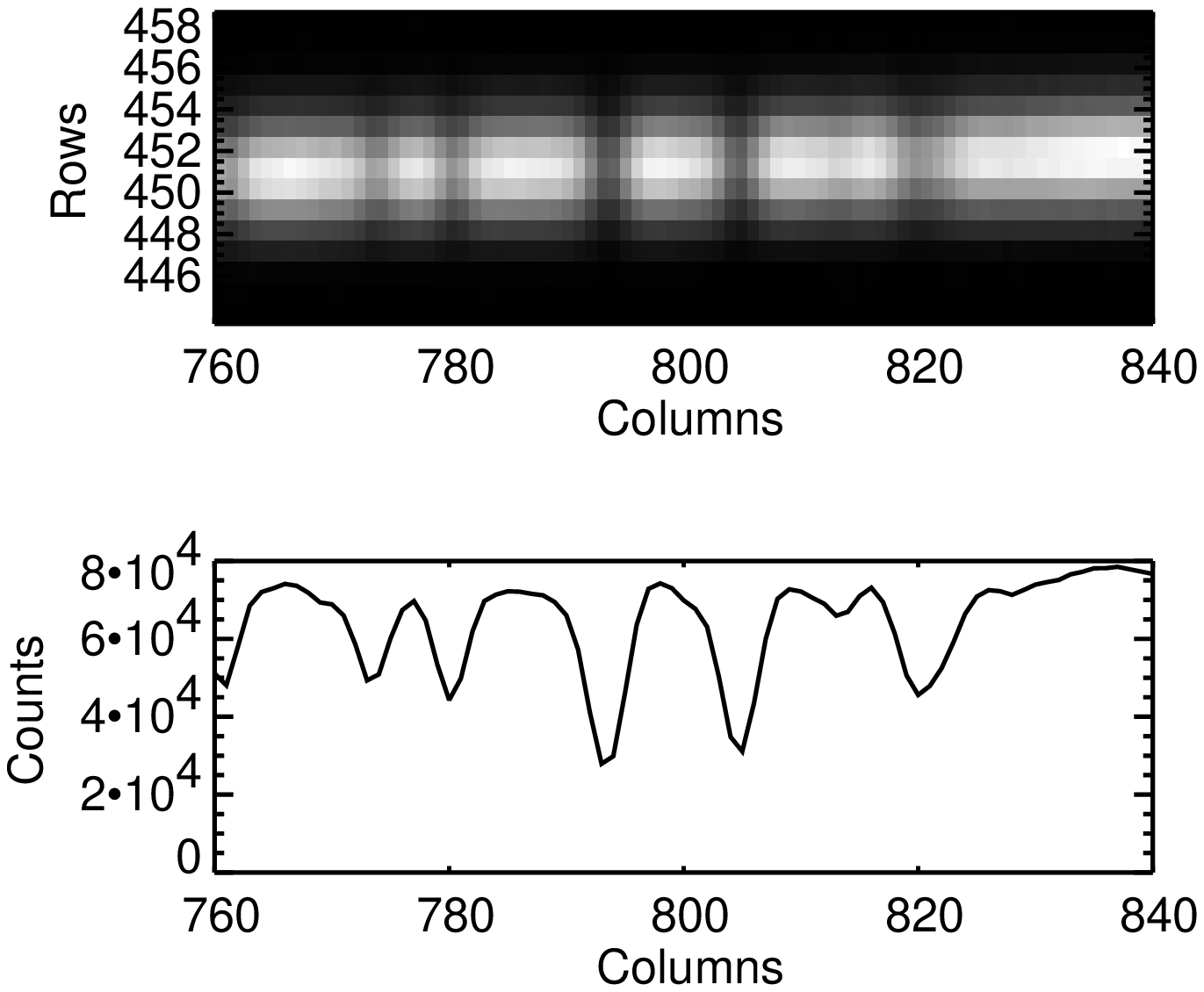}
\caption{{\it Top:} A portion of a spectral order 
from a raw CCD image of the spectrum.  Any emission line should
span the full width of the stellar spectrum, as it is caused by
seeing at the slit.
{\it Bottom:} The reduction process converts a two dimensional spectrum   
into a one dimensional spectrum.}  
\label{reduction}  
\end{figure}

The recorded two--dimensional raw spectra are reduced in the usual way  
to one--dimensional spectra.  The bias is subtracted from all images  
including both the stellar spectra and the flat--field exposures.  Each  
spectral order recorded by the CCD has a width proportional to the  
diameter of the stellar seeing disk along the length of the slit. The  
typical seeing at Keck/HIRES of 0.7--1.2 arcsec yields orders of width  
$\sim$4 CCD rows (each ``row'' being the on--chip sum of neighboring  
original CCD rows).  Cosmic rays are removed by searching for  
7--$\sigma$ sharp intensity spikes at one row compared to the  
intensity in a neighboring row (at the same wavelength).  We are  
indebted to Jason Wright for this cosmic--ray removal algorithm.  The  
stellar spectra are divided by the flat--field exposures.    
The two--dimensional spectral orders are compressed to one dimension by  
summing the recorded counts within the $\sim$10 rows  
centered on each order.  We thank Eric Nielsen for the careful reduction of
CCD images. 
Figure \ref{reduction} shows the results  
of this process. The signal--to--noise ratio per pixel in orders 72  
through 89 of a typical 1-D spectrum is $\sim$200.  
  
\subsection{Overview of the Detection Scheme}  
   
We search for laser emission lines in the reduced spectra by comparing
each spectrum to a reference spectrum of the same star, searching for
statistically significant differences.  Over time, a star's spectrum
should remain nearly constant except for minor fluctuations due to
photon noise and photospheric changes.  A large difference at one
wavelength could signify the presence of a laser signal that had been
detected during one observation but not during the other.  

We would fail to detect an emission line if it occured at the
same wavelength in both the observed and reference spectra.
A laser fired continuously from the rest frame 
of the stellar photosphere would go undetected.  However,
such a continuous laser would be detected in both spectra at different
wavelengths if the laser's velocity vector were to change between
observations, e.g. due to orbital motion around its host star.  The resulting
Doppler shift permit detections in the difference spectrum.
  
For each star, a series of algorithms is applied to all of its spectra.
First a ``reference'' spectrum is chosen.  Then, one by one, the other
``test'' spectra are aligned in wavelength with the reference for
comparison (see \S 3.2).  Candidate laser lines are found in the test spectra by
subtracting the reference spectrum and searching for positive
differences, or ``spikes.''  One test spectrum is then used as a
secondary reference to check for spikes in the primary reference
spectrum.  The spikes are subjected to a selection criterion based on
their width, to ensure that their shapes are consistent with the
instrumental profile of HIRES (see \S 3.3).  Spikes that are narrower than the
instrumental profile are likely to be caused by cosmic rays hitting
the CCD that eluded detection in the CCD reduction process described
in \S 2.1.  

A monochromatic laser line would have an intrinsic width narrower than
the resolution of the spectrometer.  However, as the laser light
passes through the HIRES optical system, it would be smeared out by
the HIRES point spread function (PSF), giving the detected emission
line an apparent distribution in wavelength characteristic of the
HIRES PSF.  Thus candidate laser lines can be distinguished from most
cosmic rays, most CCD flaws, and stellar emission lines by their
predictable spread in wavelength that would be merely the HIRES PSF,
as described in detail by Valenti et al. (1995).  Notably, emission
lines having widths that are narrower than possible from a star (with
its thermal, collisional, turbulent, and rotational broadening
mechanisms) would stand out as possibly ``nonastrophysical'' in
origin.  All of the resulting candidate emission lines of any width,
excluding the apparent cosmic rays hits, from the spectra of each star
are stored for further analysis as described in \S 4.  In sections \S
3.1, \S 3.2, and \S 3.3, we describe in detail the processes by which
we searched for emission lines.
  
\subsection{Detection Limits for Laser Emission}  

To be detected, a laser signal must compete against the light of its host star.
A laser beam carries three   advantages over starlight.  First, the
laser light is concentrated in a narrow cone of tiny solid
angle relative to  starlight, which is omnidirectional.  Second, a
monochromatic laser   line falls into one resolution element,
with a width of $\sim$2.3 pixels, while starlight, composed of
a wide range ($\sim$4000 \AA) of wavelengths, is spread out
into roughly 50,000 resolution elements.  Lastly, the laser 
needs only to outshine the photon noise of the stellar flux, not the
flux itself, within that resolution element since we are looking
at differences in the spectra.    To estimate these advantages
of the laser signal, we consider the   solid angles of the starlight
and a diffraction--limited laser beam.     
\begin{eqnarray}  
  \Omega_{\rm star}&=& 4\pi \\  
  \Omega_{\rm laser}&\approx& \Bigg({\lambda \over a}\Bigg)^2  
\label{omegas}  
\end{eqnarray}   
  
\noindent  
Here $\lambda$ is the wavelength of the laser light and $a$ is the  
aperture size of the laser transmitter.  The fraction of starlight contained  
within the laser beam is then given by  
  
\begin{equation}  
f_{\rm star}={1 \over 4\pi} \Bigg({\lambda \over a}\Bigg)^2    
\label{frac1}  
\end{equation}  
  
We note that the intrinsic width of a laser line will be much   
narrower than the HIRES resolution of $\sim$0.075~\AA,
thereby placing most all of the laser light into a single  
resolution element.  
The stellar spectral energy distribution   
may be approximated by a blackbody   
curve.   For a Solar--type star, the SED has a width of   
$\sim$4000~\AA, yielding the number of resolution elements, $N_{\rm star}$, into  
which the stellar luminosity falls:  
  
\begin{equation}  
N_{\rm star}\approx{4000\mbox{ \AA} \over 0.075\mbox{ \AA}}=5.3 \times 10^4  
\label{nstar}  
\end{equation}  
  
\noindent  
The fraction of starlight contained within one   
resolution element   
is therefore given by  
  
\begin{equation}  
f_{\rm star, \mbox{ }\Delta\lambda}={1 \over 4\pi}   
\Bigg({\lambda \over a}\Bigg)^2 {1 \over 5.3 \times 10^4}    
\label{frac2}  
\end{equation}  
  
The laser light needs only to outshine the stellar photon noise since we  
are detecting differences in a star's spectrum over time.  
The photon noise relative to the flux itself, $\sigma$, for one spectrum is given by  
1/(S/N) where S/N is the signal to noise ratio per resolution element.   
Since the laser must compete with the photon noise of two spectra in one  
resolution element, we have  
  
\begin{equation}  
\sigma \approx {\sqrt{2} \over \mbox{S/N}}  
\label{noise1}  
\end{equation}  
  
Finally we find that the laser power necessary for an $n\sigma$   
detection is  
  
\begin{equation}  
P_{\rm laser}={1 \over 4\pi} \Bigg({\lambda \over a}\Bigg)^2   
{1 \over 5.3 \times 10^4} {n \sqrt{2} \over \mbox{S/N}} L_{\rm star}  
\label{power1}  
\end{equation}  
  
\noindent  
where $L_{\rm star}$ is the luminosity of the star.  As a representative case,   
we take $\lambda$=4500~\AA, S/N=200, n=6, $L_{\rm star}=L_{\rm sun}=
3.8~\times~10^{26}$ W, and assume Keck to Keck communication 
so $a$=10 m, which gives     
  
\begin{equation}  
P_{\rm laser} = 50 \mbox{ kW}  
\label{power2}  
\end{equation}  
  
\noindent  
Thus for this representative case, one can detect a 50 kW laser.  
  
The previous calculation assumes that   
the laser is on continuously.  If, however,  
a nanosecond pulsed laser signal were transmitted, Equations \ref{power1}  
and \ref{power2} would need to be multiplied by the factor  
${\Delta t_{\rm exp} \over {N_{\rm pulses} \times 10^{-9}\mbox{ s}}}$  
where $\Delta t_{\rm exp}$ is the exposure time of the observation and $N_{\rm pulses}$  
is the number of pulses received in that time.  Using a typical exposure time  
of 10 minutes and assuming 1 pulse is received, the minimum necessary power of a  
detectable nanosecond pulsed laser is $P_{\rm pulsed}=3 \times 10^{16}$ W.    
Thus, the minimum necessary energy of
a detectable nanosecond pulsed laser is 30 MJ.

For comparison, the most powerful continuous laser on Earth is the Free Electron
Laser (FEL) developed at Jefferson Laboratory in Virginia.  The FEL has 
reached an average power of 1.7 kW and the laser is expected to achieve 
average powers up to $\sim$10 kW with an upgrade (Jefferson Lab 2001).  
Lawrence Livermore National Laboratory produced the world's most powerful 
pulsed laser, the Petawatt laser.  During its 440 femtosecond pulse, it 
delivered a power greater than 10$^{15}$ W (LLNL 2001). 
  
We can also determine a minimum detectable flux, $F_{\rm laser}$,
  
\begin{equation}  
F_{\rm laser} = {P_{\rm laser} \over A_{\rm laser}}  
\end{equation}  
  
\noindent  
where $A_{\rm laser}$ is the area of the laser beam at the Earth   
and given by  
  
\begin{equation}  
A_{\rm laser}={\pi \over 4} \Bigg({\lambda \over a}\Bigg)^2 d^2  ,
\end{equation}  
  
\noindent   
where $d$ is the distance from the laser to Earth.  The flux limit  
can now be expressed as  
  
\begin{equation}  
F_{\rm laser} = {4 \over \pi}{1 \over 5.3 \times 10^4}{n\sqrt{2} \over \mbox{S/N}}  
{L_{\rm star} \over {4 \pi d^2}} ,
\end{equation}  
  
\noindent  
where $L_{\rm star}/ 4 \pi d^2$ is the flux of the star.  
If we take $d=100$ light  
years with the previously adopted values, the minimum flux of a detectable  
laser is  
  
\begin{equation}  
F_{\rm laser} = 3.4 \times 10^{-17} \mbox{ W/m$^2$}  
\end{equation}  
  
\noindent  
  
Thus, we can detect monochromatic laser light from nearby stars  
if the flux is at least $3.4~\times~10^{-17}$~W/m$^2$ at Earth.  
  
\section{Spectral Analysis}  
  
A detailed description of the algorithms developed to analyze  
the spectra is given in this section.  
  
\subsection{Choosing Reference Spectra}  
   Two reference spectra are required for each star.  The primary  
reference spectrum is compared to all the other test spectra.    
The second is compared to the primary reference spectrum.  Since
only   positive differences are considered, any candidate laser
lines   in the first reference would not be flagged during its
comparison   with the test spectra.  A second reference, one of the
previous   test spectra, is used to account for this
possibility.

The two reference spectra are chosen as follows.  Almost every  
star has at least one ``template'' spectrum.  In the planet--search  
project, the stellar radial velocities are measured with respect  
to the template spectra.  These observations are of good quality with
high signal--to--noise ratios.  If there are multiple--template spectra for
a    star, the two with the highest signal to noise are chosen as
the    reference spectra.  If there is only one template, it is used
as the    primary reference spectrum and another observation with
high signal    to noise is chosen as the second reference.  Some
stars which are new    to the extrasolar planet search catalog do
not yet have templates.  For    these stars, the two observations
with the highest signal to noise are    chosen as the references.
     
\subsection{Aligning Spectra}  
  
Aligning a pair of spectra precisely in both wavelength and flux is   
important.  If the spectra are exactly the same, except for minor   
fluctuations due to photon noise, the fractional   
standard deviation, $\sigma$, of the difference spectrum is expected to be
  
\begin{equation}  
\sigma=\sqrt{{1 \over { \mbox{(S/N)}}_1^2}+{1 \over { \mbox{(S/N)}}_2^2}}  
\end{equation}  
  
\noindent  
where (S/N)$_1$ and (S/N)$_2$ are the signal--to--noise ratios of   
the two spectra.  For two spectra with signal--to--noise ratios of   
300, $\sigma$ can, in principle, be as low as 0.005.  It is important  
to align the two spectra well enough such that the observed  
error is as close to the theoretical error as possible in order to   
detect the smallest differences.

We line up a pair of spectra order by order.  Various normalization  
and shifting algorithms are applied to both the reference and test  
spectra, as follows.  
The first step is to put the orders on roughly the same flux  
scale by dividing by the continuum.  The continuum is found by first  
partitioning an order into 10 bins.  Each bin is assigned a temporary  
continuum at the 85th percentile in flux.  A third--order polynomial is  
fit to these individual continua to normalize the entire spectral  
order.  

Each spectrum is Doppler--shifted to correct for the Earth's orbital
motion, which causes a maximum wavelength shift of $\sim$1 \AA~
corresponding to $\sim$30 pixels.  We determine the Doppler shift of
the test spectrum relative to the reference spectrum by minimizing
$\chi^2$, and then shift the test spectrum so that the two are
aligned.  The final shifts are precise to $\sim$0.01 pixel, and
fractional pixel displacements of the spectra are accomplished with a
spline interpolation.  This shifting is first carried out on segments
of length 280 pixels, and then it is done on a 40--pixel scale
(``chunks'').  If any spectrum contains cosmic rays (or laser emission
lines), it is difficult to align.  Such features are removed
temporarily (replaced with a spline interpolation) to carry out the
alignment.  Operating on chunks of 40 pixels allows us to account for
the wavelength dependence of the Doppler shift and to correct any
slopes in the flux with wavelength.

The resulting (temporarily spike--free) reference and test spectra  
are subtracted to yield a difference spectrum, from which  
the standard deviation is determined and recorded as the  
``noise'' for each spectral chunk.    
Finally the reference and test spectra are subtracted with all spikes  
included, leaving a difference chunk which can  
be searched for laser--line candidates.  Figure 2 shows a chunk of  
difference spectrum both with (top) and without (bottom)  
flux spikes of unidentified origin.

\subsection{ Criteria for Laser Line Candidates}

A laser beam would travel through the telescope and the HIRES  
spectrometer optics just as starlight would.  Moreover, the laser
beam   travels through the same column of Earth's atmosphere,
acquiring the same wavefront distortions and seeing profile
as does the star.   Therefore, an emission line from a laser would
span the whole width of   a spectral order (perpendicular to
dispersion) as would the stellar spectrum on the CCD.  While the
intrinsic width of the laser line in wavelength would be smaller than
the resolution of HIRES, it would be broadened by the point--spread
function of HIRES.  The B1 slit used in HIRES projects to a width of
2.0 pixels on the CCD.  The instrumental profile of HIRES from both
its optics and the slit width is known from careful measurements
(Valenti et al. 1995) and from simple Gaussian fitting of emission
lines from a thorium-argon lamp.  The instrumental profile with our
HIRES setup has a width of 2.3 pixels (FWHM).  Thus, we expect any
laser line to have an observed width of 2.3 pixels.

We search for candidate emission lines in the stellar spectra as
follows.   Each 40--pixel chunk of difference spectrum has a
corresponding   fractional photon noise, $\sigma$, associated with
it as described in   \S 3.2.  We pass through all the chunks and
note spikes having   heights above $6\sigma$.  This threshold is
chosen to exclude   essentially all noise but detect a signal from a
laser with the lowest   detectable power.    

We also require that candidate laser lines exhibit a width that
is   at least as broad as the HIRES instrumental profile, with
FWHM=2.3 pixels, allowing us to discriminate 
against   (i.e., reject) cosmic--ray
hits.  Most cosmic rays produce electrons in   a single pixel or
occasionally in two neighboring pixels, clearly   narrower than the
HIRES PSF.  To implement this ``width'' criterion,   we require that
the shorter of the two adjacent pixels on either side   of each
6--$\sigma$ spike contain at least 30\% of the photons in that  
central spike.  Examination of the known HIRES PSF and thorium  
comparison lines (see Figure \ref{thor}) shows that this width  
criterion would be met by almost any emission line that had been
broadened   by the HIRES PSF.  Any spike in the difference spectrum
that does not   meet both the 6--$\sigma$ height criterion and this
width criterion is   rejected from future consideration as they are
probably cosmic rays.  Spikes   that do meet both criteria are
stored in a database that includes the   location and
characteristics of that spike, for later statistical analysis (\S
4).  We note that emission lines having widths {\em   greater} than
the PSF will be detected and stored as candidate   signals.  Thus
our survey would detect intermittent emission lines of   any origin,
laser or astrophysical, if they are sufficiently strong.

\section{Statistical Analysis of Candidates}

The spectral analysis described in \S 3 yields a   database of
surviving candidate emission lines in each   of the $\sim$12,500
Keck spectra of our 577 target stars.  A typical star has 
$\sim$13 surviving candidate emission lines in its $\sim$20
spectra.  For each candidate, the
database contains the   star, date of observation, the wavelength,
the peak number of photons,    and the width of the candidate
line.   Detailed examination shows that most (if not all)    of
the surviving candidate   emission lines are simply cosmic ray hits
that we had failed to reject.  

To discriminate a true emission line from these cosmic rays,  
we apply a statistical analysis on the collection of emission--line   
candidates from all spectra (typically 20) for a given star.    
We search for an overabundance of  
candidates (``hits'') in the vicinity of any particular wavelength.  
In contrast, cosmic rays would hit at random locations on the CCD,  
and would not favor any particular wavelength.  
A clustering of hits near one wavelength from all the spectra of  
one star could indicate laser emission.

We note that the Earth's orbital motion will not cause a Doppler   
shift of the laser line in our analysis, as we have already forceably
shifted the reference and test stellar spectra onto the same wavelength scale (\S 3).  
However, acceleration of the emitter relative to the star   
(such as due to orbital motion) during the  
typical 4 years of observations would cause the  
laser line to be Doppler--shifted to neighboring wavelengths.

We make a grid representing the CCD with the spectral orders partitioned
into approximately 30 segments of roughly 100 pixels each.  The  
segments overlap one another by half of their size to account for the  
possibility of an abundance of hits split into two neighboring  
segments.  We record the number of emission--line candidates in each segment.

For each star, we plot the number of 100--pixel segments against the
number of hits in the segments.  Most segments have no hits at all,
and only a few have one hit.  These infrequent hits (none or one hit
out of $\sim$20 observations in a wavelength segment) are consistent
with random cosmic ray hits.  To this histogram, we fit a Poisson
distribution (see Figure \ref{hist}), and determine the probabilities
of finding an excess number of hits, $n$, in any wavelength segment.
A star exhibiting an excess number of hits in a segment, i.e. a low
probability of the events occurring randomly, would stimulate further
investigation as a potential emission line.

The fit of a Poisson distribution to the histogram of hits is done as  
follows.  We fit the low--$n$ (0, 1, 2) portion of the histogram and  
neglect the high--$n$ tail.  This approach avoids contaminating the  
Poisson mean, $\mu$, with actual emission lines in the high--$n$ tail.  
The first empty bin in the histogram, $n_0$, is determined and an  
estimate of the expected mean number of hits per segment, $\mu$, is  
calculated neglecting those segments with $n \ge n_0$.  

The Poisson probability of finding $n$ cosmic ray hits   
in any given wavelength segment is given by (Taylor 1982):  
  
\begin{equation}  
P_{\mu}(n)=e^{-\mu} {\mu^n \over n!}  
\label{poisson}  
\end{equation}  
  
\noindent  
Using the estimated $\mu$, we calculate $P_{\mu}(n)$ for $n=0, 1, ...,  
n_{0}-1$, and check if any of the probabilities are less than 1\%.  If  
so, we exclude those bins from the fit as well if $n_0>2$.  The fit   
is then applied to bins from $n=0$ to some $n_{\rm max}$ hits.  
  
The estimated mean number of hits per segment  
may not necessarily yield a Poisson distribution that gives 
the best fit to the observed histogram.    
The best Poisson fit may be better determined by finding 
the lowest $\chi^2$ given by   
  
\begin{equation}  
{\chi^2}=\sum_{n=0}^{n_{\rm max}} {(O_n - E_n)^2 \over E_n}  
\label{chi2}  
\end{equation}  
  
\noindent
where $O_n$ is the observed number of segments containing $n$  
candidate hits and $E_n$ is the expected number with $n$ hits, based  
on a Poisson distribution.  The only free parameter is $\mu$, the mean  
number of hits per segment.  We vary $\mu$ to find the value  
corresponding to the minimum of $\chi^2$.  The resulting Poisson  
probabilities are multiplied by the total number of segments and  
adopted as the best fit to the observed histogram.  
  
We now determine the probabilities of finding $n$ hits in any given  
segment, extrapolating beyond $n_{\rm max}$.  Segments having an  
excessive number of hits with probabilities less than 0.01\% are  
noted as serious candidate emission lines, as they occur in excess  
numbers above the background cosmic ray hits.  Since there are about  
500 wavelength segments on our grid of the CCD for each star, we  
expect to flag one in 20 stars spuriously, i.e. due to a chance  
coincidence of cosmic rays.  
  
We find that indeed 45 stars were flagged as having a wavelength  
segment with an excess number of hits at the 0.01\% false-alarm level.  
Each of the surviving flagged candidate laser lines, of which there  
were 187, were examined by eye in the raw CCD images.  
  
\section{An Apparent Emission--Line Detection}  
  
Our search for intermittent emission lines revealed an apparent  
detection for the star HD29528 (K0V).  Three separate spectra  
obtained on different dates contained strong candidates at the same  
wavelength.  The hits were noted in both of two overlapping spectral  
segments for this star. The probability that these three apparent  
emission lines were merely spurious cosmic rays occurring within a  
segment of length $\sim$200 km s$^{-1}$ is easily calculated from the  
Poisson occurrence of cosmic rays (\S 4).

Figure \ref{hist} shows the observed histogram of the number of  
segments versus number of hits along with the best fitting Poisson  
distribution.  The probability of randomly obtaining 3 hits in a  
segment for this star is 1.23 $\times$ 10$^{-7}$ or 1 in 8 million.

Figure \ref{las1} shows these candidate laser lines in the reduced  
spectra of the three observations.  The aligned reference and test  
spectra are shown and twice the difference spectra are below them.  
The candidate laser lines are in spectral order 83 and the central  
peaks are in pixel 1436 with respect to the reference spectrum.

We examined the raw spectra to see if the candidate laser lines span  
the whole width of the spectral order, indicating whether the  
light passed through the telescope and spectrometer.  Figure  
\ref{fakelas} shows the raw CCD image of the spectrum of HD29528  
obtained on JD 2451793.1, centered on the first candidate laser line.  
The reference spectrum is shown below it.  Beneath the reference is  
the raw difference spectrum making it easier to see the apparent laser  
line.  The other two candidate laser lines found in observations  
obtained on JD 2451882.98 and on JD 2451884.08 are not shown, but  
those apparent emission lines also span the width of the spectral  
order.  Because all three hits occur at the same pixel (and hence  
wavelength) with respect to the star's reference spectrum, and because 
they   span the entire width of the spectral order, the hits are unlikely to  
be cosmic rays.  These are surviving laser line candidates thus far in  
the analysis.

To further investigate these candidate emission lines in HD29528, we  
compared the spectrum of HD29528 to the solar spectrum using ``{\it  
The Solar Spectrum 2935~\AA~to 8770~\AA}'' (Moore \et $~$1966).  The  
rest wavelengths of the lines are 4307.31~\AA~and there are no listed  
absorption lines due to the Earth's atmosphere nor anomalous Solar  
emissions lines at this wavelength.  

We further examined the spectrum of a rapidly rotating B--type star  
taken on the same night as the reference spectrum for HD29528.  The B  
star spectrum should have no prominent spectral features.  A strange  
drop in flux of 20\% is seen in the spectrum, exactly at pixel 1436 in  
order 83, precisely where the hits were found with respect to the  
reference spectrum of HD29528 (see Figure \ref{bstar}).  This clearly  
spurious feature in the B star strongly suggests that our candidate  
laser lines are merely artifacts caused by a flaw in the CCD.  

Indeed, a close examination of the raw CCD image of the reference  
spectrum of HD29528 shows the flaw in the CCD (see Figure \ref{flaw}).  
This flaw is certainly the source of the false detection.  
  
As a side note to this false detection of an emission line, one may  
calculate the necessary power of a laser that could have produced a  
``signal'' similar to that detected.  In Eqn \ref{power1} we assume a  
transmitter aperture, $a$=10 meters.  The other variables are known.  
The wavelength of the 13$\sigma$ ``signal'' is 4307~\AA, and the  
signal--to--noise ratio of the spectra is $\sim$120.  The luminosity of a K0  
star is 0.42 $L_{\rm sun}$ (Carroll and Ostlie 1996).  Using these values  
in equations \ref{power1}, we find $P_{\rm laser}=68$ kW.  This would be the  
required power of a laser to produce the spurious signal in Figure  
\ref{fakelas}.  This test supports our expectation that lasers having  
power of 50 kW would have been detected.  
  
\section{Results}  

For each of the 45 stars that exhibited an excess of candidate  
emission lines (above the cosmic ray background noise) we examined the  
original raw CCD images from HIRES.  Approximately 200 raw spectra were  
examined to follow up on the 45 flagged stars.  
Stellar emission lines were detected in the spectra of two flare stars,
HIP5643 (M4.5) and HIP92403 (M3.5).  Both stars exhibited
varying emission of the Balmer line H$\beta$. 

Interestingly, many  
candidate emission lines were located along the steep slopes of  
absorption lines in the reduced spectra.  However, the corresponding  
emission spikes were not apparent in the raw CCD image.  Such emission  
spikes are apparently neither cosmic rays nor emission lines.  
Instead, it is likely that these spikes located in absorption lines  
are caused simply by tiny changes in the instrumental profile of HIRES  
from one observation to the next.  The changes in the instrumental  
profile could result from different atmospheric seeing of the star  
image at the slit (making the slit effectively a different size) or  
from small focus changes in HIRES itself.  The steep sides of  
absorption lines have local intensities that are very sensitive to the  
instrumental profile.  
It is also possible that the spikes located
in absorption lines are residual numerical artifacts from the
spectral alignment process since interpolation routines normally
give the largest errors on steep slopes.  
Thus these candidate emission spikes located  
within absorption lines were rejected as spurious.

All other candidate emission lines were visually inspected and found  
to be inconsistent with actual emission lines.  Some were multiple,  
surviving cosmic rays that occurred, by chance, near the same  
wavelength.  Some emission spikes were caused by bad pixels or flaws  
in the CCD.  {\em No lines consistent with laser emission
were found in any of our stars.}

\section{Discussion}

We would have detected time--varying emission lines from any of
the 577 main sequence stars above a threshold of a few percent of
the continuum flux.  Emission lines of constant 
intensity would also have been detected 
had they Doppler--shifted relative to the star 
during the $\sim$4 yr of observations.  Lines having
intrinsic widths from arbitrarily narrow to tens of \kms would have
been detected.  No narrow lines consistent with laser emission were 
detected.  However, naturally occurring emission lines were found
in the spectra of two stars.  

Emission lines in the spectra of Solar--type stars are rare between
4000~\AA~-- 5000~\AA.  Such lines can arise occasionally in the
chromospheres or coronae of the most magnetically active stars.  In
some ($\sim$20\%) F, G, K--type main sequence stars, the cores of the
Balmer absorption lines are ``filled in'' by a few percent due to
chromospheric emission (Herbig 1985, Robinson, Cram, \& Giampapa
1990).  We would have detected temporal variation in that
chromospheric filling, if it exceeded a few percent of the continuum
level.  No such variations were detected in any of our F, G, K--type
main sequence stars.  This nondetection indicates that such chromospheric
variation in the cores of the strong Balmer lines, notably H$\beta$
and H$\gamma$, did not exceed a few percent of the continuum level
in these stars.  Flare stars, most of spectral type M0 or later,
more commonly exhibit emission of H$\beta$.  Indeed, we detected
time--varying emission of H$\beta$ in two M--type flare stars.

Time--varying emission lines from stars may also arise from infalling
volatile material and comets, as is observed for $\beta$ Pic (see for
example Beust et al. 1998).  Our search would have detected such
astrophysical emission lines if they appeared or varied by more than a
few percent of the continuum flux of the star.  All such astrophysical
origins for emissions lines would yield line widths greater than our
resolution of 5 \kms due to the usual thermal and turbulent
broadening, as well as possible rotational and collisional broadening.
Hence such lines would be resolved.  Nonetheless, no emission lines due
to these effects were found in the 577 stars at our threshold of a
few percent of continuum flux.

The entrance slit of the HIRES spectrometer is rectangular with
dimensions 0.57 $\times$ 3.5 arcsec.  At the typical distance of our
stars of $\sim$50 pc, this slit corresponds to 30--175 AU at the star.
Thus, sources of emission lines residing within 30 AU of our target
stars would have been included in our spectra.  Of course, the
emission beam must be pointed toward the Keck telescope for detection.
Extinction from interstellar dust is negligible toward these target
stars due to their proximity ($d<$100 pc).

The motivation for this search stemmed from the suggestion of Schwartz
and Townes (1961) that advanced civilizations might communicate from
one colony to another within a planetary system, or from one planetary
system to another, by using optical lasers as the carrier.  The
advantages of optical communication over radio include its higher
bandwidth and its smaller diffraction beam to promote private
communication.

A nondetection of extraterrestrial intelligence carries no value
unless it excludes a plausible model of life in the universe.  One
might imagine the following model.  The Milky Way Galaxy has an age of
$\sim$10 Gyr, while only 4.6 Gyr were required to spawn a
technological species on Earth.  Approximately half of the stars in
the Galactic disk are older than the Sun, and they formed with
comparable amounts of heavy elements.  Approximately 50\% of these old
disk stars do not have a stellar companion within 100
AU (Duquennoy and Mayor 1991), thereby permitting stable planetary
orbits.  Therefore, $\sim$25\% of the stars in the Galactic disk have
requisite characteristics of adequate age, chemical composition, and
dynamical quiescence to serve as sites for the development of
technological species.  The anthropocentric nature of these stellar
criteria are apparent, and thus serve merely as a useful guide for a
model.

Coincidentally, the target stars used here are systematically
middle--aged or older and void of close companions.  They were
selected for the Doppler planet search at Keck from among stars in the
Solar neighborhood.  They were chosen to have ages greater than 2 Gyr,
as judged from the CaII H\&K chromospheric emission line, because of
the resulting photospheric stability of older stars.  Moreover the
target stars have no known stellar companions within 2 arcsec (typically
$\sim$50 AU) to prevent any companion's light from entering the
spectrometer slit (but see Vogt et al. 2002 and Nidever et al. 2002
for a few recently detected, but negligibly bright, close companions).
Moreover, spectroscopic analyses of the stellar spectra show that 90\%
of the target stars contain amounts of heavy elements that are 
within a factor of 2 of the Solar value ([Fe/H] = -0.3 to +0.3;
D.A.Fischer, private communication).

Thus, the majority of the 577 target stars for this SETI search are
older than the Sun, have roughly Solar chemical composition, and have
no perturbing stellar companions within 50 AU.  This stellar sample is
clearly biased.  Thus our observations test a limited model of the
Galaxy in which older, single stars harbor technological
civilizations.

The speculative model posits that some fraction of these target stars
contain civilizations that sent probes to, or established colonies on,
planets, moons or other platforms either in their host 
planetary system or in other systems.  
This model of behavior, albeit anthropocentric,
stems from the tendencies of Homo Sapiens toward
exploration.  To communicate with such probes or colonies over
distances of Astronomical Units to parsecs, some fraction of these
civilizations might be modeled as using electromagnetic waves, 
including optical lasers.

We model the solid angle subtended by laser beams as follows.  Some
fraction, $f_{\rm tech}$, of the total number of target stars, $N_{\rm
stars}$=577, harbor technological civilizations.  On average each such
civilization emits some number of laser beams, N$_{\rm beams}$,
perhaps in arbitrary directions.  We consider that each beam subtends
an average solid angle $\Omega_{\rm beam}$.

The total solid angle, $\Omega_{\rm Total}$, 
subtended by all arbitrarily oriented beams is

\begin{equation}
\Omega_{\rm Total} = N_{\rm stars} f_{\rm tech} N_{\rm beams} \Omega_{\rm beam}
\end{equation}

\noindent
We ignore here details of overlap of the beams, their actual distribution
of power and beam size, and any purposeful aiming toward or away from us.
Any properties of the lasers that are constructed with
the Earth as a consideration could drastically change the
probability of interception of such beams.

In our survey, we have monitored $N_{\rm stars}$= 577 stars. 
The nondetection obtained in the survey here suggests that the
total solid angle subtended by lasers emitting 
50 kW or more is constrained as 
$\Omega_{\rm Total} < 4\pi$ sr .

Thus for arbitrarily directed beams, we  
constrain the model as
  
\begin{equation}
f_{\rm tech} N_{\rm beams} \Omega_{\rm beam} < 4\pi / 577 
\end{equation}
  
The nondetection reported here suggests lasers emitting
detectable power must satisfy the above constraint for the
combination of $f_{\rm tech}$,
$N_{\rm lasers}$, $\Omega_{\rm beam}$.  
Apparently, advanced civilizations
communicating by kilowatt optical lasers are not so common as to fill 4$\pi$ sr.

As a touchstone, we consider the case of the 
diffraction--limited laser with an aperture of 10-m (Keck--to--Keck),
giving a solid angle for one beam, $\Omega_{\rm beam} \approx 10^{-15}$ sr.  
For transmitters of such aperture, a laser emitting 50 kW would be detectable.
For this case, each star would have to emit $\sim10^{13}$ arbitrarily
oriented beams on average, each with power $P \ge$ 50 kW, in order 
that one would likely be oriented, serendipitously, toward us.

Clearly, the observed nondetection implies that such numerous,
narrow beams, however implausible, do not exist in sufficient numbers
to permit likely detection.  These observations
offer a poor constraint on lasers of arbitrary orientation, as their
solid angles are simply too small in this model.  One might consider
a Galactic model consisting of wider laser beams, enhancing the
chances of interception.  In that case,
the present nondetection would impose a constraint
of fewer lasers emanating from the target stars, but would
require proportionally higher power per laser for detection
at 6--$\sigma$ above the stellar continuum.  

More meaningfully, the nondetection reported here suggests that none of
the 577 target stars harbors a civilization that has purposefully
directed a laser with $P \ge $50 kW toward Earth (for the case of 
a 10--meter diffraction--limited transmitting aperture).  This modest detection
threshold of 50 kW highlights the ease with which an advanced
civilization could signal us, if desired.  But such is apparently not
the case.

%As an example, we might consider \Omega_{\rm beam} to be that given by  
%optical light from a diffraction--limited 10-m aperture transmitter.  

\section{Acknowledgements}

We thank the SETI Institute for generous funding of this work.  We
wish to thank R.Paul Butler and Steve Vogt for their   endless hours
in obtaining the Keck/HIRES spectra.   S.Vogt's design of HIRES made
this work possible.   We thank Jason Wright for the new CCD
reduction algorithm of Keck spectra that efficiently removed cosmic
ray hits.  We thank Eric Nielsen for reducing the 12,500 CCD images
to one--dimensional spectra.  We thank Debra Fischer for chemical--abundance 
measurements of the target stars.  We also thank D.Nidever,
D.Werthimer, J.Tarter, F.Drake, M.Lampton, B.Walp and referee, William D. Cochran 
for many useful comments.   This work 
was carried out by A.R. as partial fulfillment of the of the
requirements toward a Master's Degree in physics at San Francisco
 State University.

\clearpage  
  
\begin{figure}  
\plotone{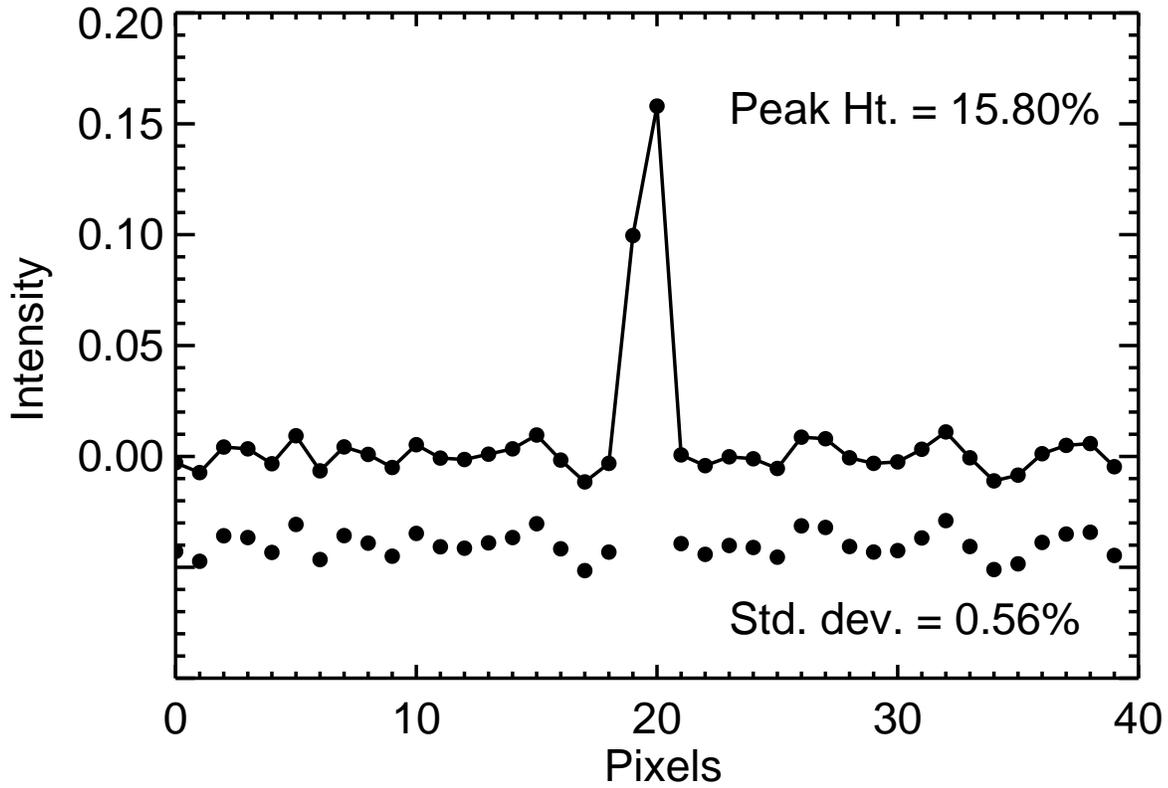}  
\caption{{\it Top:} A full chunk of difference spectrum, including spikes, which  
will be searched for laser line candidates.  The peak height is 15.80\% above
the continuum.  {\it Bottom:} The same chunk  
of difference spectrum with spikes removed and offset for clarity.  The standard deviation of  
this spike-free chunk is the reported sigma.}  
\label{diff}  
\end{figure}  
  
\begin{figure}  
\plotone{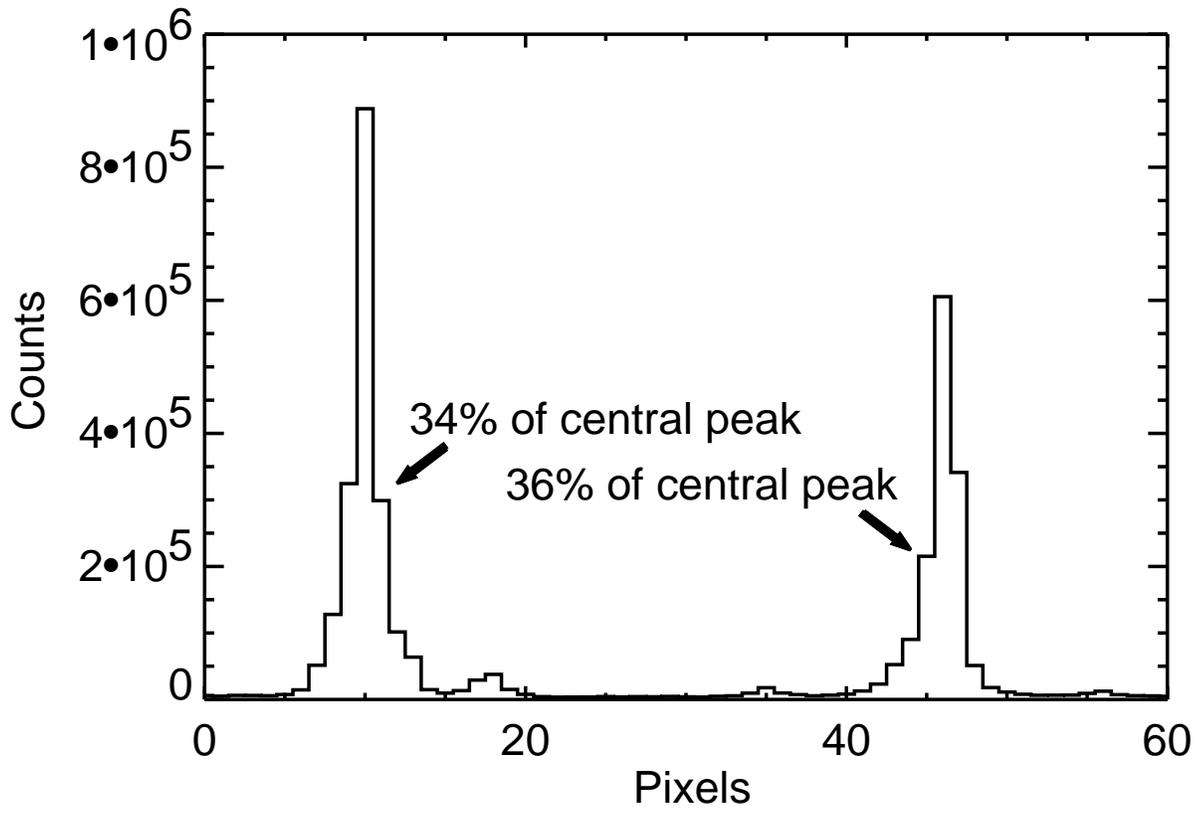}  
\caption{Emission lines in a thorium-argon spectrum showing the   
spectrometer PSF.}  
\label{thor}  
\end{figure}  
  
\begin{figure}  
\plotone{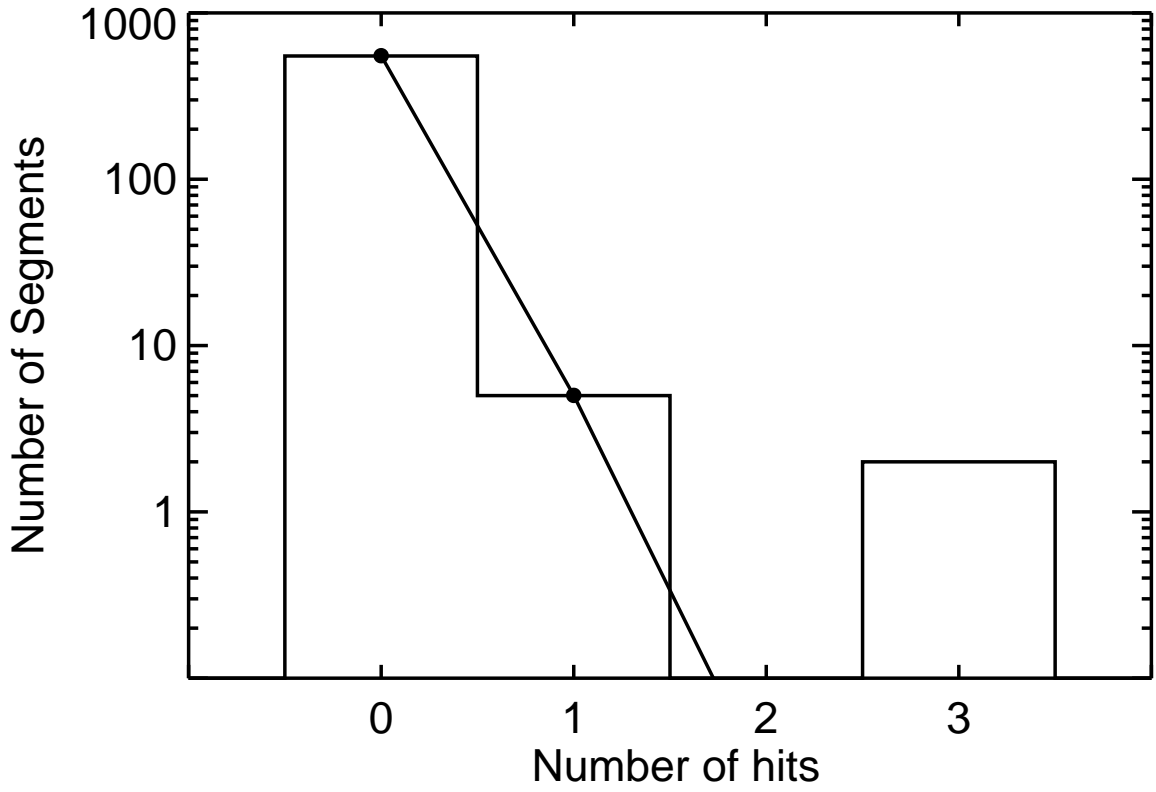}  
\caption{The observed histogram of the number of segments  
versus number of hits for the star HD29528.  The dots indicate the best   
fitting Poisson distribution with the mean number of hits per  
segment equal to 0.00907.}  
\label{hist}  
\end{figure}   
  
\begin{figure}  
%\psfig{figure=fig5.ps,width=\textwidth}
%%\centerline{\scalebox{.8}{\includegraphics{fig5.ps}}}  
\plotone{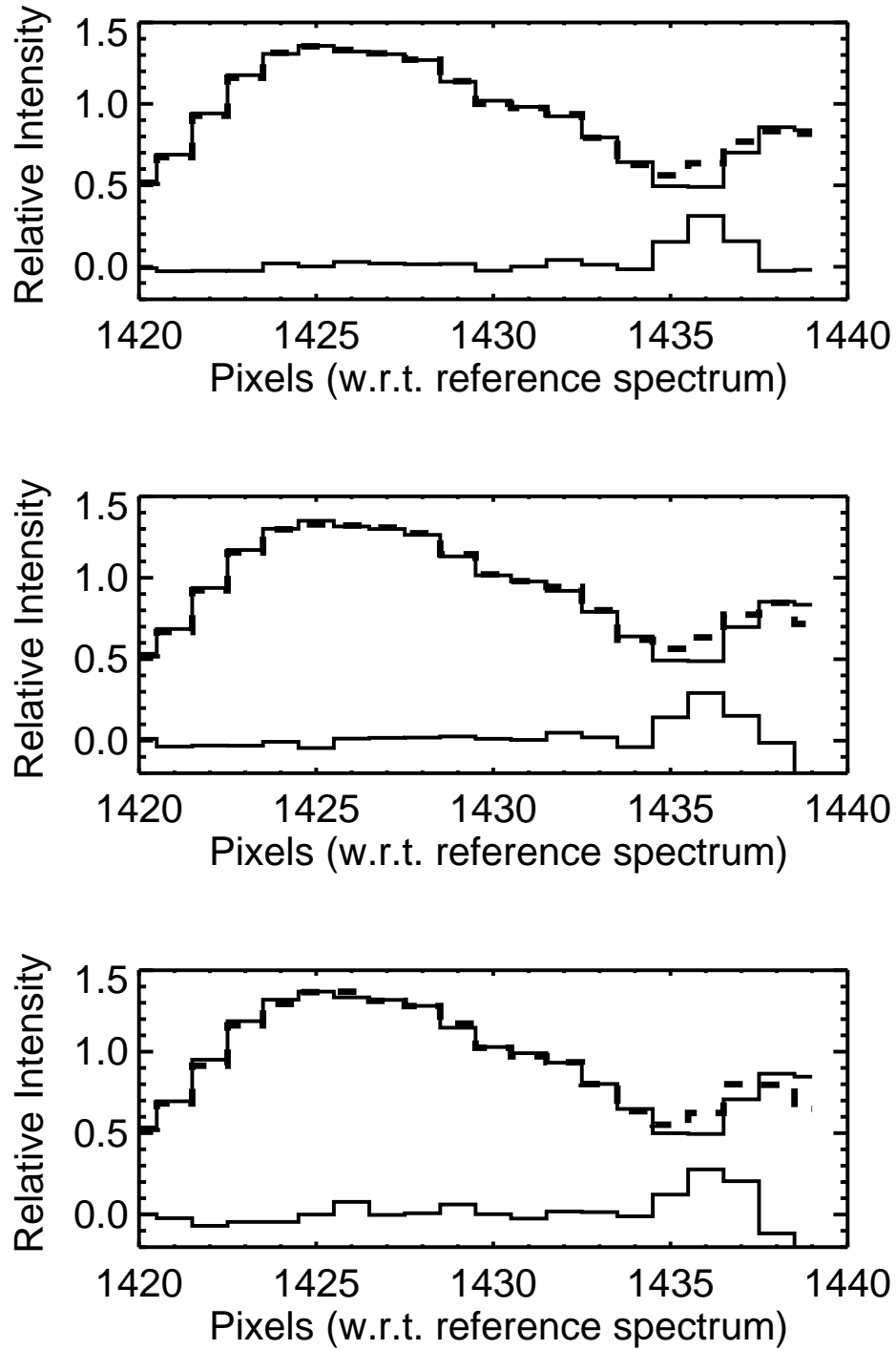}
\caption{Three hits in the reduced spectra of HD29528.  The test spectra  
(dotted lines) have been aligned with the reference spectra (solid lines).  
Twice the difference spectra are shown below.    
The candidate laser lines are in order 83 and the  
central peak is in pixel 1436 with respect to the reference spectrum.}  
\label{las1}  
\end{figure}  
  
\begin{figure}  
%\centerline{\scalebox{.35}{\rotatebox{90}{\includegraphics{fig6a.ps}}}}  
%\centerline{\scalebox{.35}{\rotatebox{90}{\includegraphics{fig6b.ps}}}}  
%\centerline{\scalebox{.35}{\rotatebox{90}{\includegraphics{fig6c.ps}}}}  
%\plotone{fig6a.ps}
%\plotone{fig6b.ps}
%\plotone{fig6c.ps}
\psfig{figure=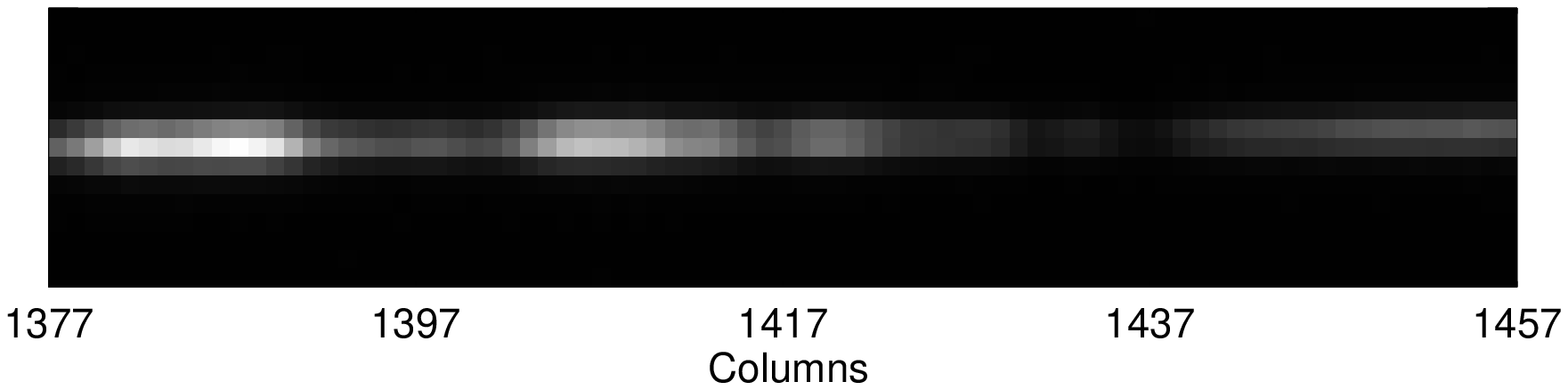,angle=90,width=\textwidth}
\psfig{figure=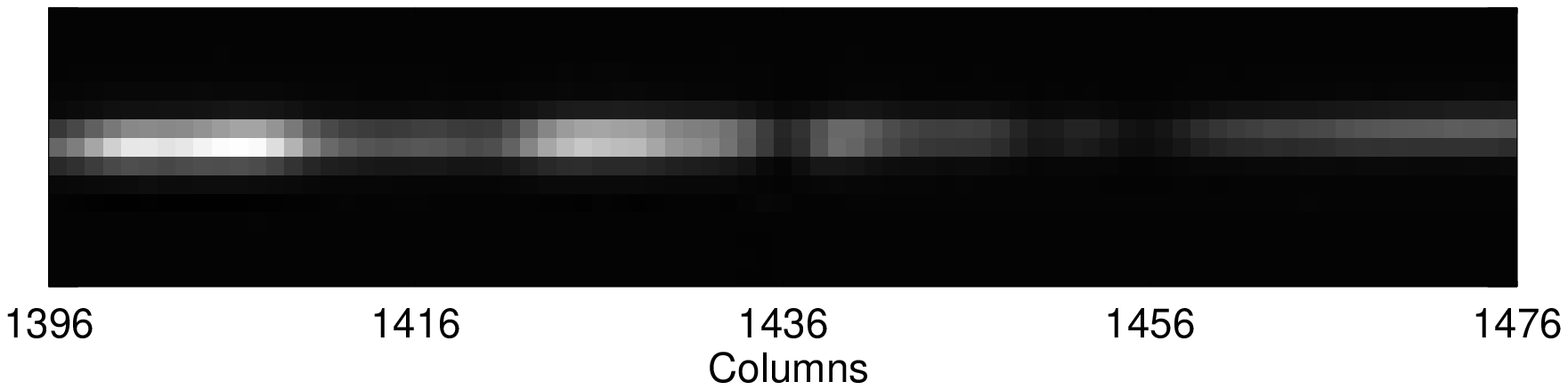,angle=90,width=\textwidth}
\psfig{figure=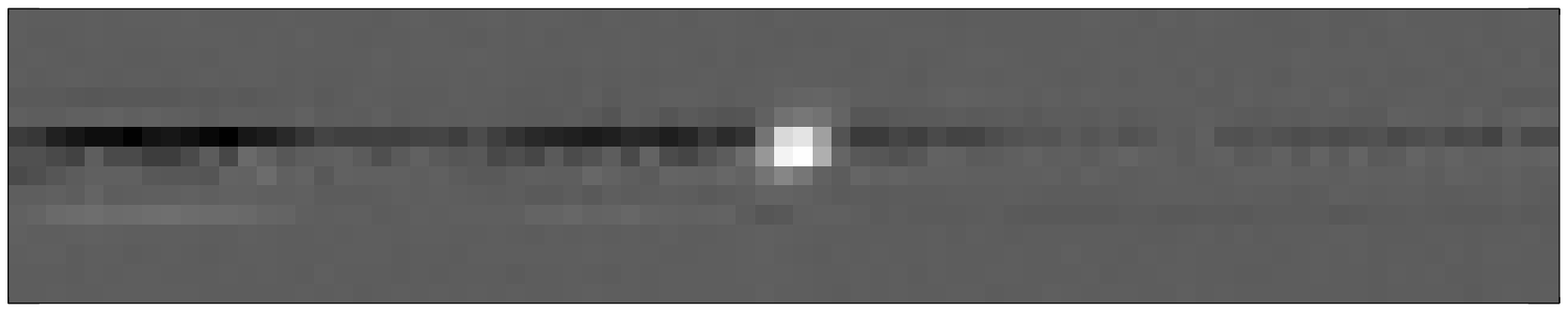,angle=90,width=\textwidth}
%\centerline{\scalebox{.8}{\rotatebox{90}{\includegraphics{fig6.ps}}}}  
\caption{An apparent narrow emission line. {\it Top:}  The raw 
CCD image of an observation of HD29528 taken
on JD 2451793.1, centered  
on the first candidate laser line. {\it Middle:} The raw CCD image  
of the reference spectrum.  {\it Bottom:} The difference of the two  
raw spectra.  A bright line spanning the width of the spectral order   
can easily be seen.}  
\label{fakelas}  
\end{figure}  
  
\begin{figure}  
\plotone{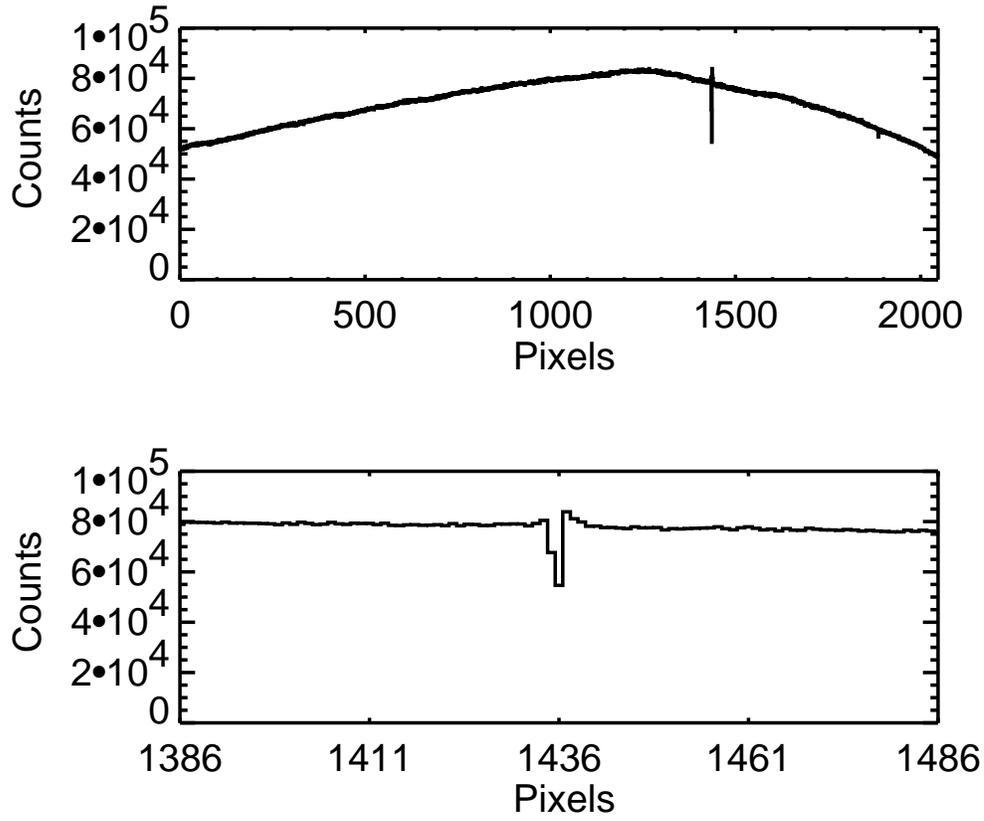}  
\caption{{\it Top:} Order 83 of a B star spectrum.  A strong feature can  
be seen at pixel 1436. {\it Bottom:} Zooming in on the bad pixels.}  
\label{bstar}  
\end{figure}   
  
\begin{figure}  
%\centerline{\scalebox{.35}{\rotatebox{90}{\includegraphics{fig8a.ps}}}}  
%\centerline{\scalebox{.35}{\rotatebox{90}{\includegraphics{fig8b.ps}}}}  
%\plotone{fig8a.ps}
%\plotone{fig8b.ps}
\psfig{figure=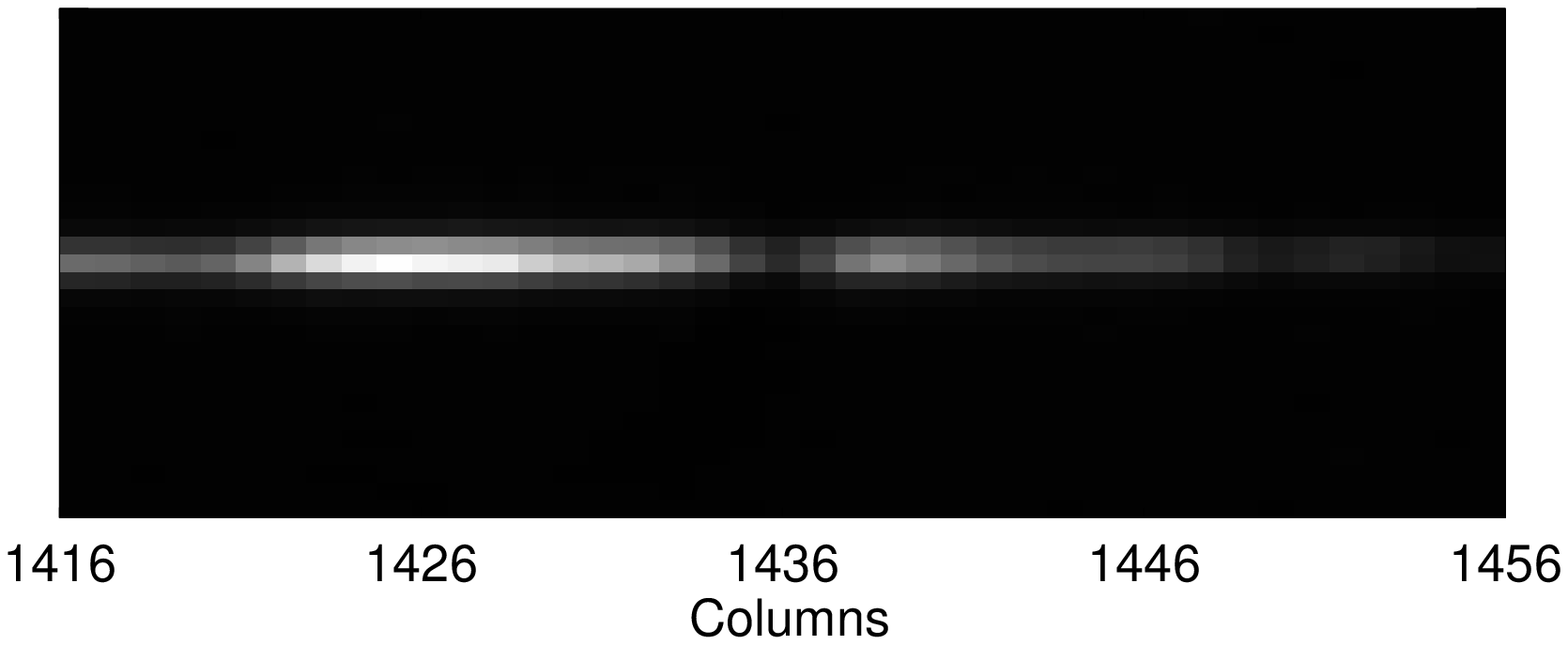,angle=90,width=\textwidth}
\psfig{figure=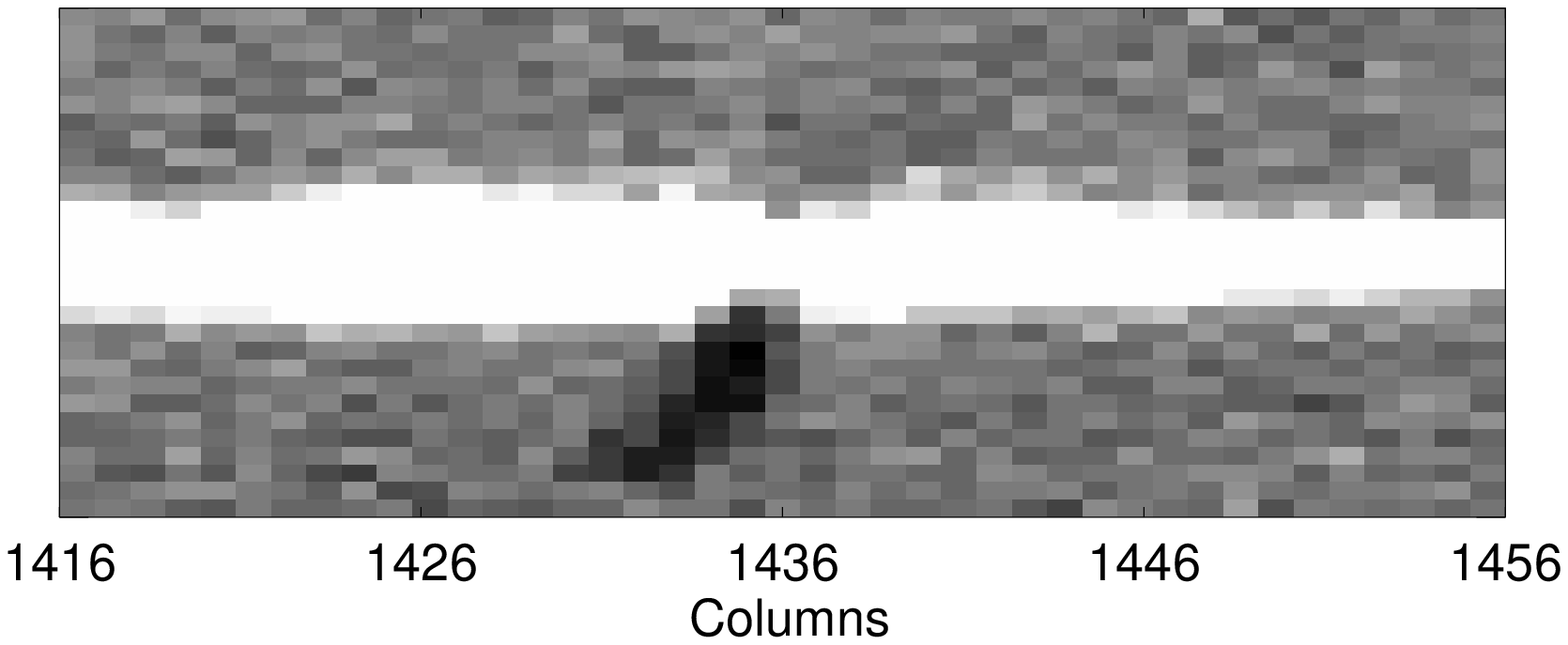,angle=90,width=\textwidth}
%\centerline{\scalebox{.8}{\rotatebox{90}{\includegraphics{fig8.ps}}}}  
\caption{A flaw in the CCD masquerades as an emission line. {\it Top:} A chunk of 
raw spectrum from order 83 of the reference  
spectrum for HD29528 centered on the location of the candidate laser lines.   
{\it Bottom:} The same chunk of spectrum with the intensity levels adjusted.   
A dark CCD flaw can be seen under the bright spectral order.}  
\label{flaw}  
\end{figure}  
 
\end{document}